\documentclass[prd,email,twocolumn,showpacs,showkeys,preprintnumbers,amsmath,amssymb,nofootinbib]{revtex4-1}
\usepackage{amsmath}
\usepackage{latexsym}
\usepackage[pdftex]{graphicx}
\usepackage{graphicx}
\usepackage{amsfonts}
\usepackage{amssymb}
\usepackage{amsmath}
\usepackage[all]{xy}
\begin{document}

\title{Some aspects of a Chern-Simons-like coupling in an external magnetic field}

\author{Patricio Gaete$^1$}
\email{patricio.gaete@usm.cl}
\author{Jos\'{e} A. Hela\"{y}el-Neto$^2$}
\email{helayel@cbpf.br}
\author{Euro Spallucci$^3$}
\email{spallucci@ts.infn.it}

\affiliation{${}^{1}$Departmento de F\'{\i}sica and Centro Cient\'{\i}fico-Tecnol\'{o}gico
de Valpara\'{\i}so, \\
Universidad T\'ecnica Federico Santa Mar\'{\i}a, Valpara\'{\i}so, Chile \\
${}^{2}$ Centro Brasileiro de Pesquisas F\'{\i}sicas,\\
Rua Xavier Sigaud, 150, Urca, 22290-180, Rio de Janeiro, Brazil\\
${}^{3}$ Dipartimento di Fisica Teorica, Universit\`a di Trieste
and INFN, Sezione di Trieste, Italy \\
\today\\}

\pacs{11.15.-q; 11.10.Ef; 11.30.Cp}

\keywords{gauge invariance, confinement, static potential}

\begin{abstract}
\noindent For a gauge theory which includes a light massive vector
field interacting with the familiar photon $U(1)_{QED}$ via a
Chern-Simons- like coupling, we study the static quantum potential.
Our analysis is based on the gauge-invariant, but path-dependent, variables formalism. The result is that the theory describes
an exactly screening phase. Interestingly enough, this result displays
a marked departure of a qualitative nature from the axionic
elctrodynamics result. However, the present result is analogous
to that encountered in the coupling between the familiar photon
$U(1)_{QED}$ and a second massive gauge field living in the
so-called hidden-sector $U(1)_h$, inside a superconducting box.
\end{abstract}

\maketitle

\pagestyle{myheadings} \markright{{\it Some aspects of a
Chern-Simons-like coupling in an external magnetic field}}

\section{Introduction}
\renewcommand{\theequation}{1.\arabic{equation}}
\setcounter{equation}{0}

Nowadays, one of the most actively pursued areas of research in
physics consists of the investigation of extensions of the Standard
Model (SM) such as axion-like particles and light extra
hidden $U(1)$ gauge bosons, in order to explain cosmological and astrophysical results. These hidden $U(1)$ gauge bosons are frequently encountered in string theories physics. Also, this subject has had a revival after recent results of the PVLAS collaboration \cite{Zavattini,Cameron,Chen,Zavattini2,Robilliard,Chou,Ahlers,Ahlers2,Popov,
Popov2,Chelouche,Gninenko,Jaeckel}. Nevertheless, although none of these searches ultimately has yielded a positive signal, the arguments in favor of the existence of axion-like particles or light extra hidden $U(1)$ gauge bosons, remain as cogent as ever.

In this perspective, it is useful to recall that the axion-like
scenario can be qualitatively understood by the existence of light
pseudoscalar bosons $\phi$ ("axions"), with a coupling to two
photons. In other terms, the interaction term in the effective
Lagrangian has the form $\mathcal{L}_I  =  - \frac{1}{4}F_{\mu \nu }
\tilde F^{\mu \nu } \phi$, where $\tilde F^{\mu \nu }  =
\frac{1}{2}\varepsilon _{\mu \nu \lambda \rho} F^{\lambda \rho }$.
However, the crucial feature of axionic electrodynamics is the mass
generation due to the breaking of rotational invariance induced by a
classical background configuration of the gauge field strength
\cite{Spallucci}, which leads to confining potentials in the presence
of nontrivial constant expectation values for the gauge field
strength $F_{\mu \nu}$ \cite{GaeteGuen}. In fact, in the case of a
constant electric field strength expectation value the static
potential remains Coulombic, while in the case of a constant
magnetic field strength expectation value the potential energy is
the sum of a Yukawa and a linear potential, leading to the
confinement of static charges. Also it is important to point out
that the magnetic character of the field strength expectation value
needed to obtain confinement is in agreement  with the current
chromo-magnetic picture of the $QCD$ vacuum \cite{Savvidy}. In
addition, similar results have been obtained in the context of the
dual Ginzburg-Landau theory \cite{Suganuma}, as well as for a theory
of antisymmetric tensor fields that results from the condensation of
topological defects as a consequence of the Julia-Toulouse mechanism
\cite{GaeteW}. 

In a general perspective, we draw attention to the fact that much of this work has been inspired by studies coming from the realms of string theory \cite{Jaeckel2,Batell} and quantum field theory \cite{Holdom,Nath,Nath2,Castelo,Masso,Foot, Singleton,Langacker}. Indeed, as was observed in \cite{Singleton2}, the introduction of a second gauge field in addition to the usual photon was pioneered in Ref. \cite{Cabibbo}, in the context of electrodynamics in the presence of magnetic monopoles \cite{Dirac}. Whereas the quantization for a system with two photons was later carried out in \cite{Hagen}. The possible existence of massive vector fields was also proposed in \cite{Okun}. A kinetic term between the familiar photon $U(1)_{QED}$ and a second gauge field has also been considered in order to explain recent unexpected observations in high energy astrophysics \cite{Nima}.

We further note that recently another possible candidate for extensions of the SM  has been studied \cite{Antoniadis}. It is the so-called Chern-Simons-like coupling scenario, which includes a light massive vector field interacting with the familiar photon $U(1)_{QED}$ via a Chern-Simons-like
coupling. As a result, it was argued that this new model reproduces the effects of rotation of the polarization plane.

Given the relevance of these studies, it is of interest to improve
our understanding of the physical consequences presented by this new
scenario (Chern-Simons-like coupling scenario). Of special interest
will be to study the connection or equivalence with the  axion-like particles and light extra hidden $U(1)$ gauge bosons scenarios. Thus, our
purpose here is to further explore the impact of a light massive
vector field in the Chern-Simons-like coupling scenario on physical
observables. To this end, we will study the screening and confinement issue. This issue is generally not discussed.  Our calculation
is accomplished by making use of the gauge-invariant but
path-dependent variables formalism along the lines of Ref.
\cite{GaeteS,GaeteJPA, GaeteSpa09,GaeteHel09}, which is a
physically-based alternative to the usual Wilson loop approach.
As we shall see, in the case of a constant magnetic field the
theory describes an exactly screening phase. This then implies
that the static potential profile obtained from both the Chern-Simons-like coupling and axionic electrodynamics models are quite different. This means that the two theories are not equivalent.  As it was shown in
\cite{GaeteGuen}, axionic electrodynamics has a different structure
which is reflected in a confining piece, which is not present in the
Chern-Simons-like coupling scenario. Incidentally, the
above static potential profile (Chern-Simons-like coupling scenario)  is similar to that encountered in the coupling between the familiar massless electromagnetism $U(1)_{QED}$ and a hidden-sector $U(1)_h$ inside a superconducting box \cite{Gaete0410}. In this way one obtains a new connection among different models describing the same physical phenomena. In this connection, the benefit of considering the present approach is to provide unifications among different models, as well as exploiting the equivalence in explicit calculations, as we shall see in the course of the discussion.

\section{Interaction energy}\label{s2}
\renewcommand{\theequation}{2.\arabic{equation}}
\setcounter{equation}{0}

We now discuss the interaction energy between static point-like
sources for the model under consideration. To carry out such study, we will
compute the expectation value of the energy operator $H$ in the
physical state $|\Phi\rangle$ describing the sources, which we will
denote by $ {\langle H\rangle}_\Phi$.

As anticipated above, the gauge theory we are considering describes the
interaction between the familiar massive photon $U(1)_{QED}$ with a
light massive vector field via a Chern-Simons- like coupling. In
this case the corresponding theory is governed by the Lagrangian
density \cite{Antoniadis}:
\begin{eqnarray}
\mathcal{L} &=&  - \frac{1}{4}F_{\mu \nu }^2 (A) - \frac{1}{4}F_{\mu
\nu }^2 (B) + \frac{{m_\gamma ^2 }}{2}A_\mu ^2  + \frac{{m_B^2
}}{2}B_\mu ^2 \nonumber\\
&-& \frac{{\kappa }}{2}\varepsilon ^{\mu
\nu \lambda \rho } A_\mu  B_\nu F_{\lambda \rho } (A),
\label{Csmag05}
\end{eqnarray}
where $m_\gamma$ is the mass of the photon, and $m_{B}$ represents
the mass for the gauge boson $B$. In particular, this alternative
theory exhibits an effective mass for the component of the photon
along the direction of the external magnetic field, exactly as it
happens with axionic electrodynamics. If we consider the model in
the limit of a very heavy $B$-field ($m_{B} \gg m_\gamma$) and we
are bound to energies much below $m_{B}$, we are allowed to
integrate over $B_\mu$ and to speak about an effective model for the
$A_\mu$-field. Then, the first crucial point is that by eliminating
the gauge field $B_\mu$ in terms of $A_\mu$ in the original
Lagrangian (\ref{Csmag05}) one gets an effective Lagrangian
$\mathcal{L}_{eff}$. This is accomplished by making use of the
following shifting for the $B_\mu$-field:
\begin{equation}
B_\mu   \equiv \tilde B_\mu   + \frac{\kappa }{2}\frac{1} {{\left(
{\Delta  + m_B^2 } \right)}}\left[ {\eta _{\mu \nu } +
\frac{{\partial _\mu  \partial _\nu  }}{{m_B^2 }}}
\right]\varepsilon ^{\mu \lambda \rho \nu } A_\mu  F_{\lambda \rho }
\left( A \right). \label{Csmag10}
\end{equation}

Once this is done, and dropping the constant factor which emerges by
integrating out the $\tilde B_\mu$-field, we arrive at the following
effective Lagrangian density
\begin{eqnarray}
\mathcal{L}_{eff}  &=&  - \frac{1}{4}F_{\mu \nu }^2  +
\frac{{m_\gamma ^2 }} {2}A_\mu  A^\mu   + \frac{{\kappa ^2
}}{4}A_\alpha  F_{\beta \gamma } \frac{1}
{{\left( {\Delta  + m_B^2 } \right)}}A^\alpha  F^{\beta \gamma } \nonumber\\
&+& \frac{{\kappa ^2 }}{2}A_\alpha  F_{\beta \gamma }
\frac{1}{{\left( {\Delta  + m_B^2 }
\right)}}A^\gamma  F^{\alpha \beta } \nonumber\\
&+& \frac{{\kappa ^2 }}{8}\tilde F^{\alpha \beta } F_{\alpha \beta }
\frac{1} {{m_B^2 \left( {\Delta  + m_B^2 } \right)}}\tilde F^{\gamma
\delta } F_{\gamma \delta }, \label{Csmag15}
\end{eqnarray}
where ${\widetilde F}_{\mu \nu }  \equiv {\raise0.7ex\hbox{$1$}
\!\mathord{\left/{\vphantom {1 2}}\right.\kern-\nulldelimiterspace}
\!\lower0.7ex\hbox{$2$}}\varepsilon _{\mu \nu \lambda \rho }
F^{\lambda \rho }$. The same result can be obtained by integrating
out the $B$ field in a path integral formulation of the model. The
integral is gaussian in $B$ and can be exactly computed leading to
the effective Lagrangian (\ref{Csmag15}). 

Before going ahead, we would like to note that from Eq. (\ref{Csmag15}) the  gauge invariance is broken and one could argue about the possibility of getting a gauge invariant result for the static potential between test charges from (\ref{Csmag15}). There are at least two available options to solve this apparent inconsistency. One way is to restore gauge invariance
by inserting Stuckelberg compensating fields either into
(\ref{Csmag05}) or into (\ref{Csmag15}). Once the compensators are
integrated out the resulting model is explicitly gauge invariant.
Unfortunately, this procedure introduces non-local effective
interaction terms which are difficult to handle. In alternative we
shall follow the Hamiltonian formulation discussed below.

As a second point, if we wish to study quantum properties of the
electromagnetic field in the presence of external electric and
magnetic fields, we should split the $A_\mu$-field as the sum of a
classical background, $\langle A_\mu \rangle$, and a small quantum
fluctuation, $a_\mu$,
\begin{equation}
A_\mu = \langle A_\mu \rangle + a_\mu. \label{Csmag20}
\end{equation}
Therefore the previous Lagrangian density, up to quadratic terms in
the fluctuations, is also expressed as 
\begin{widetext}
\begin{eqnarray}
{\cal L}_{eff}  &=&  - \frac{1}{4}f_{\mu \nu } \Omega f^{\mu \nu} + \frac{1}{2}a_\mu  M^2 a^\mu -\frac{{\kappa ^2
}}{2}f_{\gamma \beta }\left\langle {A^\gamma  } \right\rangle
\frac{{1}}{{\left( {\Delta  + m_B^2 } \right)}}
\left\langle {A_\alpha  } \right\rangle f^{\alpha \beta } + \frac{{\kappa ^2 }}{8}f_{\mu \nu } v^{\mu \nu } \frac{1} {{m_B^2
\left( {\Delta  + m_B^2 } \right)}}v^{\lambda \rho } f_{\lambda
\rho} \nonumber\\
&-& \kappa ^2 \left( {\varepsilon ^{jk0i} v_{0i} \left\langle {A_j }
\right\rangle a^m \frac{1}{{\left( {\Delta  + m_B^2 }
\right)}}f_{km} }
\right) + \kappa ^2 \left( {\varepsilon ^{jk0i} v_{0i} \left\langle {A^m }
\right\rangle a_k \frac{1}{{\left( {\Delta  + m_B^2 }
\right)}}f_{jm} }
\right) \nonumber\\
&-& \frac{{\kappa ^2 }}{2}\left( {\varepsilon ^{jk0i} v_{0i}
\left\langle {A^l } \right\rangle a_l \frac{1}{{\left( {\Delta  +
m_B^2 } \right)}}f_{jk} } \right) ,  \label{Csmag25}
\end{eqnarray} 
\end{widetext}
where $f_{\mu \nu } = \partial _\mu  a_\nu   - \partial _\nu a_\mu$,
and $\Delta \equiv \partial_\mu\partial^\mu$. $\Omega  \equiv 1 -
\kappa ^2 \frac{{\left\langle {A^i } \right\rangle \left\langle {A_i
} \right\rangle }}{{\left( {\Delta  + m_B^2 } \right)}}$, and $M^2
\equiv m_\gamma ^2  + \frac{{\kappa ^2 }}{2}\frac{{v_{0i} v^{0i}
}}{{\left( {\Delta  + m_B^2 } \right)}}$. In the above Lagrangian we
have considered the $v^{0i}\neq0$ and $v^{ij}=0$ case (referred to
as the magnetic one in what follows), and simplified our notation by
setting $\varepsilon ^{\mu \nu \alpha \beta } \left\langle{F_{\mu
\nu } } \right\rangle  \equiv v^{\alpha \beta }$. As a consequence,
the Lagrangian (\ref{Csmag25}) becomes a Maxwell-Proca-like theory
with a manifestly Lorentz violating term. 

This effective theory provide us with a suitable starting point to
study the interaction energy. However, before proceeding with the
determination of the energy, it is necessary to restore the gauge
invariance in (\ref{Csmag25}). For this end we now carry out a
Hamiltonian analysis. The canonically conjugate momenta are found to
be $\Pi ^\mu   =  - \left[ {1 - \kappa ^2 \frac{{\left\langle {A^k }
\right\rangle \left\langle {A_k  } \right\rangle }}{{\left( {\Delta
+ m_B^2 } \right)}}} \right]f^{0\mu }  + \kappa ^2
\frac{{\left\langle {A^\mu  } \right\rangle \left\langle {A_i }
\right\rangle }}{{\left( {\Delta  + m_B^2 } \right)}}f^{i0}  -
\frac{{\kappa ^2 }}{{4m_B^2 }}v^{\mu 0} \frac{1}{{\left( {\Delta  +
m_B^2 } \right)}}v^{0i} f_{0i}$. This yields the usual primary
constraint  $\Pi^{0}=0$, while the momenta are $\Pi^i  = - \left[ {1
- \kappa ^2 \frac{{\left\langle {A^k  } \right\rangle \left\langle
{A_k  } \right\rangle }}{{\left( {\Delta  + m_B^2 } \right)}}}
\right]f^{0i}  + \kappa ^2 \frac{{\left\langle {A^i} \right\rangle
\left\langle {A_k } \right\rangle }}{{\left( {\Delta  + m_B^2 }
\right)}}f^{k0}  - \frac{{\kappa ^2 }}{{4m_B^2 }}v^{i0}
\frac{1}{{\left( {\Delta  + m_B^2 } \right)}}v^{0k} f_{0k}$. This
leads us to the canonical Hamiltonian, 
\begin{widetext}
\begin{eqnarray}
 H_C  &=&  - \int {d^3 x} a_0 \left( {\partial _i \Pi ^i  + \frac{1}{2}\left\{ {m_\gamma ^2  - \frac{{{\raise0.7ex\hbox{${\kappa ^2 {\bf v}^2 }$} \!\mathord{\left/
 {\vphantom {{\kappa ^2 v^2 } 2}}\right.\kern-\nulldelimiterspace}
\!\lower0.7ex\hbox{$2$}}}}{{\left( {\Delta  + m_B^2 } \right)}}} \right\}a^0 } \right) + \int {d^3 x} \left( { - \frac{1}{2}a_i \left\{ {m_\gamma ^2  - \frac{{{\raise0.7ex\hbox{${\kappa ^2 {\bf v}^2 }$} \!\mathord{\left/
 {\vphantom {{\kappa ^2 v^2 } 2}}\right.\kern-\nulldelimiterspace}
\!\lower0.7ex\hbox{$2$}}}}{{\left( {\Delta  + m_B^2 } \right)}}} \right\}a^i } \right) \nonumber\\
&+& \int {d^3 x} \left( {\frac{1}{2}\Pi _i \left\{ {1 - \frac{{\kappa ^2 \left\langle {\bf A} \right\rangle ^2 }}{{\left( {\Delta  + m_B^2 } \right)}}} \right\}\Pi ^i } \right) + \int {d^3 x} \left( {\frac{{\kappa ^2 }}{2}\frac{1}{{\left( {\Delta  + m_B^2 } \right)}}\left( {\left\langle {\bf A} \right\rangle \cdot {\bf \Pi }} \right)^2 } \right) \nonumber\\
&+& \int {d^3 x} \left( {\frac{1}{2}B_i \left\{ {1 + \frac{{\kappa ^2 \left\langle {\bf A} \right\rangle ^2 }}{{\left( {\Delta  + m_B^2 } \right)}}} \right\}B^i } \right) + \int {d^3 x} \left\{ { - \frac{1}{2}\frac{{\kappa ^2 }}{{\left( {\Delta  + m_B^2 } \right)}}\left( {\varepsilon _{ikj} \left\langle {A^k } \right\rangle B^j } \right)^2 } \right\} \nonumber\\
&+& \int {d^3 x} \left( {\kappa ^2 \varepsilon ^{jk0i} v_{0i} \left\langle {A_j } \right\rangle a^m \frac{1}{{\left( {\Delta  + m_B^2 } \right)}}f_{km} } \right) - \int {d^3 x} \left( {\kappa ^2 \varepsilon ^{jk0i} v_{0i} \left\langle {A^m } \right\rangle a_k \frac{1}{{\left( {\Delta  + m_B^2 } \right)}}f_{jm} } \right) \nonumber\\
&+& \int {d^3 x} \left( {\frac{{\kappa ^2 }}{2}\varepsilon ^{jk0i} v_{0i} \left\langle {A^l } \right\rangle a_l \frac{1}{{\left( {\Delta  + m_B^2 } \right)}}f_{jk} } \right) ,  \label{Csmag30}
\end{eqnarray}
\end{widetext} 
where ${\bf v}$ stands for the external magnetic field ($v^{0i}$)
and $B^i$ is now the magnetic field associated to the fluctuation,
namely, $B^i\equiv\epsilon^{ijk} f_{jk}$.

Time conservation of the primary constraint $ \Pi_0=0$ yields the
following secondary constraint $\Gamma \left( x \right) \equiv
\partial _i \Pi ^i  + \left( {m_\gamma ^2  - \frac{{\kappa ^2 {\bf v}^2 }}
{2}\frac{1}{{\left( {\Delta  + m_B^2 } \right)}}} \right)a^0  = 0$.
Notice that the nonvanishing bracket $\left\{ {\Pi ^0 ,\partial _i
\Pi ^i  + \left( {m_\gamma ^2  - \frac{{\kappa ^2 {\bf v}^2
}}{2}\frac{1}{{\left( {\Delta  + m_B^2 } \right)}}} \right)a^0 }
\right\}$ shows that the above pair of constraints are second class
constraints, as expected for a theory with an explicit mass term
which breaks the gauge invariance. To convert the second class
system into first class we enlarge the original phase space by
introducing a canonical pair of fields $\theta$ and $\Pi _\theta$
\cite{GaeteSpa09}. It follows, therefore, that a new set of first
class constraints can be defined in this extended space:
\begin{equation}
\Lambda _1  = \Pi _0  + \left( {m_\gamma ^2  - \frac{{\kappa ^2 {\bf
v}^2 }}{2}\frac{1}{{\left( {\Delta  + m_B^2 } \right)}}} \right)a^0,
\label{Csmag35a}
\end{equation}
and
\begin{equation}
\Lambda _2  \equiv \Gamma  + \Pi _\theta. \label{Csmag35b}
\end{equation}

In this way the gauge symmetry of the theory under consideration has
been restored. Then, the new effective Lagrangian, after integrating
out the $\theta$-field, becomes
\begin{widetext}
\begin{eqnarray}
\mathcal{L}_{eff}  &=&  - \frac{1}{4}f_{\mu \nu } \left[
{\frac{{\left( {\Delta ^2 + a^2 \Delta  + b^2 } \right)}}{{\Delta
\left( {\Delta  + m_B^2 } \right)}}} \right]f^{\mu \nu }
-\left\langle {A^i } \right\rangle f_{i0} \frac{{1}}
{{\left( {\Delta + m_B^2 } \right)}} \left\langle {A_k }
\right\rangle f^{k0}
- \frac{{\kappa ^2 }}{2}f_{ki} \left\langle {A^k } \right\rangle
\frac{{1}} {{\left( {\Delta  + m_B^2 } \right)}} \left\langle {A_l }
\right\rangle f^{li} \nonumber\\
&+& \frac{{\kappa ^2 }}{8}v^{0i}
f_{0i} \frac{1}{{m_B^2 \left( {\Delta  + m_B^2 } \right)}}v^{0k}
f_{0k}, \label{Csmag40}
\end{eqnarray}
\end{widetext}
where $a^2  \equiv m_B^2  + m_\gamma ^2 \left( {1 - \kappa ^2
\frac{{\left\langle {A_k  } \right\rangle \left\langle {A^k  }
\right\rangle }} {{m_\gamma ^2 }}} \right)$, and $ b^2  = m_\gamma
^2 \left( {m_B^2  - \frac{{\kappa ^2 {\bf v}^2 }}{{2m_\gamma ^2 }}}
 \right)$. We observe that to get the above theory we have ignored the last three
terms in (\ref{Csmag30}) because it add nothing to the static
potential calculation, as we will show it below. In other words, the
new effective action (\ref{Csmag30}) provide us with a suitable
starting point to study the interaction energy without loss of
physical content.

We now turn our attention to the calculation of the interaction
energy. In order to obtain the corresponding Hamiltonian, the
canonical quantization of this theory from the Hamiltonian analysis
point of view is straightforward and follows closely that of our
previous work \cite{GaeteSpa09,GaeteHel09}. The canonical momenta read  $\Pi ^\mu   =  - \left( {\frac{{\Delta
^2  + a^2 \Delta  + b^2 }} {{\Delta \left( {\Delta  + m_B^2 }
\right)}}} \right)f^{0\mu } + \kappa ^2 \frac{{\left\langle {A^\mu
} \right\rangle \left\langle {A_k } \right\rangle }}{{\left( {\Delta
+ m_B^2 } \right)}}f^{k0} \\ + \frac{{\kappa ^2 }}{4}\frac{{v^{0\mu }
}}{{m_B^2 }}\frac{1} {{\left( {\Delta  + m_B^2 } \right)}}v^{0k}
f_{0k}$, and one immediately identifies the usual primary constraint
$\Pi ^i  =  - \left( {\frac{{\Delta ^2  + a^2 \Delta  + b^2 }}
{{\Delta \left( {\Delta  + m_B^2 } \right)}}} \right)f^{0i} + \kappa
^2 \frac{{\left\langle {A^i } \right\rangle \left\langle {A_k }
\right\rangle }}{{\left( {\Delta  + m_B^2 } \right)}}f^{k0} +
\frac{{\kappa ^2 }}{4}\frac{{v^{0i} }}{{m_B^2 }}\frac{1} {{\left(
{\Delta  + m_B^2 } \right)}}v^{0k} f_{0k}$. The canonical
Hamiltonian is thus given by
\begin{widetext}
\begin{eqnarray}
H_C  &=& \int {d^3 x} \left\{ { - a_0 \partial _i \Pi ^i  +
\frac{1}{2}B^i \frac{{\left( {\Delta ^2  + a^2 \Delta + b^2 }
\right)}}{{\Delta \left( {\Delta  + m_B^2 } \right)}}B^i } \right\}
- \frac{1}{2}\int {d^3 } x\Pi _i \frac{{\left( {\Delta  + m_B^2 }
\right)}}
{{\left( {\Delta  + a^2  + \frac{{b^2 }}{\Delta }} \right)}} \Pi^i \nonumber\\
&+& \frac{{\kappa ^2 }}{2}\int {d^3 x} \Pi _i \left\langle {A^i }
\right\rangle \frac{{\left( {\Delta  + m_B^2 } \right)}}{{\left(
{\Delta  + a^2  + \frac{{b^2 }}{\Delta }} \right)^2}}\left\langle
{A^k } \right\rangle \Pi _k
+ \frac{{\kappa ^2 }}{2}\int {d^3 x} f_{ki} \left\langle {A^k }
\right\rangle \frac{1}{{\left( {\Delta  + m_B^2 }
\right)}}\left\langle {A_l } \right\rangle f^{li}
\nonumber\\
&+& \frac{{\kappa ^2 {\bf v}^2 }}{{8m_B^2 }}\int {d^3 x} \Pi _i
\frac{{\left( {\Delta + m_B^2 } \right)}}{{\left( {\Delta  + a^2  +
\frac{{b^2 }}{\Delta }} \right)^2 }}\Pi ^i  ,\label{Csmag45}
\end{eqnarray}
\end{widetext}
where $a^2  = m_B^2  + m_\gamma ^2  + \kappa ^2 \left\langle {\bf A}
\right\rangle ^2$ and $b^2  = m_\gamma ^2 m_B^2  + \frac{{\kappa ^2
}}{2}{\bf v}^2$. Since our energies are all much below $m_B$, it is
consistent with our considerations to neglect $\kappa ^2
\left\langle {\bf A} \right\rangle ^2$ with respect to $m_B^2$. This
implies that $a^2$ and $b^2$ should be taken as: $a^2  = m_B^2$ and
$b^2  = m_\gamma ^2 m_B^2  + \frac{{\kappa ^2 }}{2}{\bf v}^2$.

The consistency condition $\dot \Pi _0  = 0$ leads to the usual
Gauss constraint $\Gamma_1 \left( x \right) \equiv \partial _i \Pi
^i=0$. It is also possible to verify that no further constraints are
generated by this theory. Consequently, the extended Hamiltonian
that generates translations in time then reads $H = H_C + \int
{d^2}x\left( {c_0 \left( x \right)\Pi _0 \left( x \right)
 + c_1 \left(x\right)\Gamma _1 \left( x \right)} \right)$, where $c_0 \left(
x\right)$ and $c_1 \left( x \right)$ are arbitrary Lagrange
multipliers. Moreover, it follows from this Hamiltonian that
$\dot{a}_0 \left( x \right)= \left[ {a_0 \left( x \right),H} \right]
= c_0 \left( x \right)$, which is completely arbitrary. Since $
\Pi^0 = 0$ always, neither $a^0$ and $\Pi^0$ are of interest in
describing the system and may be discarded from the theory. 
If a new arbitrary coefficient $c(x) = c_1 (x) - A_0 (x)$ is introduced the Hamiltonian may be rewritten as 
\begin{widetext}
\begin{eqnarray}
H  &=& \int {d^3 x} \left\{ {c(x)\partial _i \Pi ^i  + \frac{1}{2}B^i \frac{{\left( {\Delta ^2  + a^2 \Delta  + b^2 } \right)}}{{\Delta \left( {\Delta  + m_B^2 } \right)}}B^i } \right\} - \frac{1}{2}\int {d^3 } x\Pi _i \frac{{\left( {\Delta  + m_B^2 } \right)}}{{\left( {\Delta  + a^2  + \frac{{b^2 }}{\Delta }} \right)}}\Pi^i \nonumber\\
&+& \frac{{\kappa ^2 }}{2}\int {d^3 x} \Pi _i \left\langle {A^i } \right\rangle \frac{{\left( {\Delta  + m_B^2 } \right)}}{{\left( {\Delta  + a^2  + \frac{{b^2 }}{\Delta }} \right)^2}}\left\langle {A^k } \right\rangle \Pi _k + \frac{{\kappa ^2 }}{2}\int {d^3 x} f_{ki} \left\langle {A^k } \right\rangle \frac{1}{{\left( {\Delta  + m_B^2 } \right)}}\left\langle {A_l } \right\rangle f^{li}  \nonumber\\
&+& \frac{{\kappa ^2 {\bf v}^2 }}{{8m_B^2 }}\int {d^3 x} \Pi _i
\frac{{\left( {\Delta  + m_B^2 } \right)}}{{\left( {\Delta  + a^2  +
\frac{{b^2 }}{\Delta }} \right)^2 }}\Pi ^i. \label{Csmag50}
\end{eqnarray}
\end{widetext}

We can at this stage impose a subsidiary on the vector potential
such that the full set of constraints become second class. A
particularly convenient choice is found to be
\begin{equation}
\Gamma _2 \left( x \right) \equiv \int\limits_{C_{\xi x} } {dz^\nu }
a_\nu \left( z \right) \equiv \int\limits_0^1 {d\lambda x^i } a_i
\left( {\lambda x} \right) = 0,     \label{Csmag55}
\end{equation}
where  $\lambda$ $(0\leq \lambda\leq1)$ is the parameter describing
the spacelike straight path $ x^i = \xi ^i  + \lambda \left( {x -
\xi } \right)^i $, and $ \xi $ is a fixed point (reference point).
There is no essential loss of generality if we restrict our
considerations to $ \xi ^i=0 $. The choice (\ref{Csmag55}) leads to
the Poincar\'e gauge \cite{Gaete99}. As a consequence, we can now
write down the only nonvanishing Dirac bracket for the canonical
variables
\begin{eqnarray}
\left\{ {a_i \left( x \right),\Pi ^j \left( y \right)} \right\}^ *
&=&\delta{ _i^j} \delta ^{\left( 2 \right)} \left( {x - y} \right)
\nonumber\\
&-&
\partial _i^x \int\limits_0^1 {d\lambda x^j } \delta ^{\left( 2
\right)} \left( {\lambda x - y} \right). \label{Csmag60}
\end{eqnarray}

At this point, we have all the elements necessary to find  the
interaction energy between point-like sources for the model under
consideration. As we have already indicated, we will calculate the
expectation value of the energy operator $H$ in the physical state
$|\Phi\rangle$. In this context, we recall that the physical state
$|\Phi\rangle$ can be written as
\begin{eqnarray}
\left| \Phi  \right\rangle & \equiv &  \left| {\overline \Psi
\left(
\bf y \right)\Psi \left( {\bf 0} \right)} \right\rangle \nonumber\\
&=& \overline \psi \left( \bf y \right)\exp \left(
{iq\int\limits_{{\bf 0}}^{\bf y} {dz^i } a_i \left( z \right)}
\right)\psi \left({\bf 0} \right)\left| 0 \right\rangle,
\label{Csmag65}
\end{eqnarray}
where $\left| 0 \right\rangle$ is the physical vacuum state. The
line integral is along a spacelike path starting at $\bf 0$ and
ending at $\bf y$, on a fixed time slice. Notice that the charged
matter field together with the electromagnetic cloud (dressing)
which surrounds it, is given by $\Psi \left( {\bf y} \right) = \exp
\left( { - iq\int_{C_{{\bf \xi} {\bf y}} } {dz^\mu A_\mu  (z)} }
\right)\psi ({\bf y})$. Thanks to our path choice, this physical
fermion then becomes $\Psi \left( {\bf y} \right) = \exp \left( { -
iq\int_{\bf 0}^{\bf y} {dz^i  } A_{i} (z)} \right)\times \psi ({\bf
y})$. In other terms, each of the states ($\left| \Phi
\right\rangle$) represents a fermion-antifermion pair surrounded by
a cloud of gauge fields to maintain gauge invariance.

Further, by taking into account the structure of the Hamiltonian
above, we observe that 
\begin{eqnarray}
\Pi _i \left( x \right)\left| {\overline \Psi  \left( \bf y
\right)\Psi \left( {{\bf y}^ \prime  } \right)} \right\rangle  &=&
\overline \Psi  \left( \bf y \right)\Psi \left( {{\bf y}^ \prime }
\right)\Pi _i \left( x \right)\left| 0 \right\rangle \nonumber\\
&+& q\int_ {\bf y}^{{\bf y}^ \prime  } {dz_i \delta ^{\left( 3
\right)} \left( {\bf z - \bf x} \right)} \left| \Phi
\right\rangle.\nonumber\\
\label{Csmag65b}
\end{eqnarray}
Having made this observation and since the fermions are taken to be
infinitely massive (static sources), we can substitute $\Delta$ by
$-\nabla^{2}$ in Eq. (\ref{Csmag50}). Therefore, the expectation
value $\left\langle H \right\rangle _\Phi$ is expressed as
\begin{equation}
\left\langle H \right\rangle _\Phi   = \left\langle H \right\rangle
_0 + \left\langle H \right\rangle _\Phi ^{\left( 1 \right)} +
\left\langle H \right\rangle _\Phi ^{\left( 2 \right)},
\label{Csmag70}
\end{equation}
where $\left\langle H \right\rangle _0  = \left\langle 0
\right|H\left| 0 \right\rangle$. The $\left\langle H \right\rangle
_\Phi ^{\left( 1 \right)}$ and $\left\langle H \right\rangle _\Phi
^{\left( 2 \right)}$ terms are given by
\begin{widetext}
\begin{eqnarray}
\left\langle H \right\rangle _\Phi ^{\left( 1 \right)}  &=&  -
\frac{{b^2 B}}{2} \int {d^3 x} \left\langle \Phi  \right|\Pi _i
\frac{{\nabla ^2 }}
{{\left( {\nabla ^2  - M_1^2 } \right)}}\Pi ^i \left| \Phi  \right\rangle +\frac{{b^2 B}}{2}\int {d^3 x} \left\langle \Phi  \right|\Pi _i
\frac{{M_2^2 }} {{M_1^2 }}\frac{{\nabla ^2 }}{{\left( {\nabla ^2  -
M_2^2 } \right)}} \Pi ^i \left| \Phi  \right\rangle, \nonumber\\
\label{Csmag75a}
\end{eqnarray}
and
\begin{eqnarray}
\left\langle H \right\rangle _\Phi ^{\left( 2 \right)}  &=& m_B^2
b^2 B\int {d^3 x} \left\langle \Phi  \right|\Pi _i \frac{1}{{\left(
{\nabla ^2  - M_1^2 } \right)}} \Pi ^i \left| \Phi  \right\rangle - m_B^2 b^2 B\int {d^3 x} \left\langle \Phi  \right|\Pi _i
\frac{{M_2^2 }} {{M_1^2 }}\frac{1}{{\left( {\nabla ^2  - M_2^2 }
\right)}}\Pi ^i \left|
\Phi  \right\rangle,  \nonumber\\
\label{Csmag75b}
\end{eqnarray}
\end{widetext}
where $B = \frac{1}{{M_2^2 \left( {M_2^2  - M_1^2 } \right)}}$,
$M_1^2  \equiv \frac{{a^2 }}{2}\left[ {1 + \sqrt {1 - \frac{{4b^2 }}
{{a^4 }}} } \right]$, and $M_2^2  \equiv \frac{{a^2 }} {2}\left[ {1
- \sqrt {1 - \frac{{4b^2 }}{{a^4 }}} } \right]$.

We have neglected the terms in (\ref{Csmag50}) where $\left( {\Delta
+ a^2  + \frac{{b^2 }}{\Delta }} \right)^2$ appears in the
denominator, the reason being that we wish to compute an
interparticle potential, which expresses the effects of photons
exchange in the low-energy (or low-frequency) limit. Therefore,
these terms we are mentioning are suppressed in view of the presence
of higher power of the frequency in the denominator. Another important point to be highlighted in our discussion comes
from the expressions for $M_1^2$ and $M_2^2$. Our treatment is only
valid under the assumption that $a^4 > 4b^2$. However, this
condition is equivalent to taking $\kappa ^2 {\bf v}^2  <
\frac{{m_B^4 }}{2}$, which is perfectly compatible with our
approximation. So, we restrict ourselves to an external magnetic
field such that $|{\bf v}| < \frac{{m_B^2 }}{{2\kappa ^2 }}$.

Using Eq. (\ref{Csmag65b}), the $\left\langle H \right\rangle _\Phi
^ {\left( 1 \right)}$ and $\left\langle H \right\rangle _\Phi ^
{\left( 2 \right)}$ terms can be rewritten as
\begin{widetext}
\begin{eqnarray}
\left\langle H \right\rangle _\Phi ^{\left( 1 \right)}  &=&  -
\frac{{b^2 Bq^2 }}{2}\int {d^3 x} \int_{\bf y}^{{\bf y}^ \prime  }
{dz_i^ \prime  \delta ^{(3)} \left( {{\bf x} - {\bf z}^ \prime  }
\right)} \frac{{\nabla ^2 }}{{\left( {\nabla ^2  - M_1^2 } \right)}}
\times \int_{\bf y}^{{\bf y}^ \prime  } {dz^i } \delta ^{(3)}
\left( {{\bf x} - {\bf z}} \right)
\nonumber\\
&+& \frac{{b^2 Bq^2 }}{2}\frac{{M_2^2 }}{{M_1^2 }}\int {d^3 x}
\int_{\bf y}^{{\bf y}^ \prime  } {dz_i^ \prime  \delta ^{(3)} \left(
{{\bf x} - {\bf z}^ \prime  } \right)} \frac{{\nabla ^2 }} {{\left(
{\nabla ^2  - M_2^2 } \right)}} \int_{\bf y}^{{\bf y}^ \prime } {dz^i } \delta ^{(3)} \left(
{{\bf x} - {\bf z}} \right), \label{Csmag80a}
\end{eqnarray}
and
\begin{eqnarray}
\left\langle H \right\rangle _\Phi ^{\left( 2 \right)}  &=& m_B^2
b^2 Bq^2 \int {d^3 x} \int_{\bf y}^{{\bf y}^ \prime  } {dz_i^ \prime
\delta ^3 \left( {{\bf x} - {\bf z}^ \prime  } \right)}
\frac{1}{{\left( {\nabla ^2  - M_1^2 } \right)}}
\int_{\bf y}^{{\bf y}^ \prime  } {dz^i } \delta ^{(3)}
\left( {{\bf x} - {\bf z}} \right) \nonumber\\
&-& \frac{{m_B^2 b^2 Bq^2M_2^2 }}{{M_1^2 }}\int {d^3 x} \int_{\bf
y}^{{\bf y}^ \prime  } {dz_i^ \prime  \delta ^{(3)} \left( {{\bf x}
- {\bf z}^ \prime  } \right)}  \frac{1} {{\left( {\nabla ^2  - M_2^2} \right)}} 
\int_{\bf y}^{{\bf y}^ \prime  }
{dz^i } \delta ^{(3)} \left( {{\bf x} - {\bf z}} \right). \nonumber\\
\label{Csmag80b}
\end{eqnarray}
\end{widetext}
Following our earlier procedure \cite{GaeteS,GaeteHel09}, we see
that the potential for two opposite charges located at ${\bf y}$ and
${\bf y^{\prime}}$ takes the form
\begin{eqnarray}
 V &=&  - \frac{{q^2 }}{{4\pi }}\frac{1}{{a^2 \sqrt {1 - {\raise0.7ex\hbox{${4b^2 }$} \!\mathord{\left/
 {\vphantom {{4b^2 } {a^4 }}}\right.\kern-\nulldelimiterspace}
\!\lower0.7ex\hbox{${a^4 }$}}} }} \nonumber\\ 
&\times& \left[ {\left( {M_1^2  - m_B^2 } \right)\frac{{e^{ - M_1 L} }}{L} - \left( {M_2^2  - m_B^2 } \right)\frac{{e^{ - M_2 L} }}{L}} \right] . \nonumber\\ 
 \label{Csmag85}
\end{eqnarray}
Consequently, our analysis reveals that the theory under consideration describes an exactly screening phase. It is important to realize that expression (\ref{Csmag85}) displays a marked departure of a qualitative nature from the result from axionic elctrodynamics. As already mentioned, axionic electrodynamics has a different structure which is reflected in a confining piece, which is not present in the Chern-Simons-like coupling scenario. It is to be noted that the choice of the gauge is in this development really arbitrary. Put another way, being the formalism completely gauge invariant, we would obtain exactly the same result in any gauge. We also note here that by considering the limit $b \to 0$, we obtain 
a theory of two independent uncoupled  $U(1)$ gauge bosons, one of which is massless. In such a case, one can easily verify that the
static potential is a Yukawa-like correction to the usual static Coulomb
potential.

Finally, the following remark is pertinent at this point. It should be noted that  by substituting $B_\mu$ by $\partial_\mu \phi$ in (\ref{Csmag05}), the theory under consideration assumes the form \cite{Gaete06}
\begin{equation}
{\cal L} =  - \frac{1}{4}F_{\mu \nu }^2  + \frac{{m_\gamma ^2 }}{2}
+ \frac{1}{2}\partial _\mu  \phi \partial ^\mu  \phi  - \frac{\kappa
}{{2m_B }}\varepsilon ^{\mu \nu \lambda \rho } F_{\mu \nu }
F_{\lambda \rho } \phi, \label{Csmag90}
\end{equation}
which is similar to axionic electrodynamics. In fact, it is worth recalling here that axionic electrodynamics is described by \cite{GaeteGuen}
\begin{equation}
{\cal L} =  - \frac{1}{4}F_{\mu \nu } F^{\mu \nu }  -
\frac{g}{8}\phi \varepsilon ^{\mu \nu \rho \sigma } F_{\mu \nu }
F_{\rho \sigma } + \frac{1}{2}\partial _\mu  \phi \partial ^\mu \phi
- \frac{{m_A^2 }}{2}\phi ^2,  \label{Csmag90b}
\end{equation}
hence we see that both theories are quite different.

Thus, after performing the integration over $\phi$ in (\ref{Csmag90}), the
effective Lagrangian density reads
\begin{eqnarray}
{\cal L} &=&  - \frac{1}{4}F_{\mu \nu }^2  + \frac{{m_\gamma ^2
}}{2}A_\mu ^2 - \frac{{\kappa ^2 }}{{8m_B^2 }}\varepsilon ^{\mu \nu
\lambda \rho } F_{\mu \nu } F_{\lambda \rho } \frac{1}{{\nabla ^2
}} \nonumber\\
&\times&\varepsilon ^{\alpha \beta \gamma \delta } F_{\alpha \beta }
F_{\gamma \delta }. \label{Csmag95}
\end{eqnarray}
This expression can now be rewritten as
\begin{eqnarray}
{\cal L} &=&  - \frac{1}{4}f_{\mu \nu }^2  + \frac{{m_\gamma ^2
}}{2}a_\mu ^2 - \frac{{\kappa ^2 }}{{2m_B^2 }}\varepsilon ^{\mu \nu
\alpha \beta } \left\langle {F_{\mu \nu } } \right\rangle
\varepsilon ^{\lambda \rho \gamma \delta } \left\langle {F_{\lambda
\rho } } \right\rangle \nonumber\\
&\times& f_{\alpha \beta } \frac{1}{{\nabla ^2
}}f_{\gamma \delta } , \label{Csmag100}
\end{eqnarray}
where $\left\langle {F_{\mu \nu } } \right\rangle$ represents the
constant classical background. Here $f_{\mu \nu } =\partial _\mu
a_\nu -\partial _\nu a_\mu$ describes fluctuations around the
background. The above Lagrangian arose after using $ \varepsilon
^{\mu \nu \alpha \beta } \left\langle {F_{\mu \nu } } \right\rangle
\left\langle {F_{\alpha \beta } } \right\rangle=0$ (which holds for
a pure electric or a pure magnetic background). By introducing the
notation $\varepsilon ^{\mu \nu \alpha \beta }
 \left\langle{F_{\mu \nu } } \right\rangle  \equiv v^{\alpha \beta }
 $ and $\varepsilon ^{\rho \sigma \gamma \delta } \left\langle {F_{\rho
\sigma } } \right\rangle  \equiv v^{\gamma \delta }$, expression
(\ref{Csmag100}) then becomes
\begin{equation}
{\cal L} =  - \frac{1}{4}f_{\mu \nu }^2  + \frac{{m_\gamma ^2
}}{2}a_\mu ^2 - \frac{{\kappa ^2 }}{{2m_B^2 }}v^{\alpha \beta }
f_{\alpha \beta } \frac{1}{{\nabla ^2 }}v^{\gamma \delta } f_{\gamma
\delta} , \label{Csmag105}
\end{equation}
where the tensor $v^{\alpha \beta }$ is not arbitrary, but satisfies
$\varepsilon ^{\mu \nu \alpha \beta } v_{\mu \nu } v_{\alpha \beta
}=0$.

Following the same steps employed for obtaining (\ref{Csmag85}), the static potential is expressed as 
\begin{equation}
V= - \frac{{q^2 }}{{4\pi }}\frac{{e^{ - ML} }}{L}, \label{Csmag130}
\end{equation}
with $ M^2  \equiv m_\gamma ^2  + \frac{{\kappa ^2 }}{{m_B^2 }}{\bf
v}^2$ .
Again, the theory describes a screening phase, as we have just seen above. 
\section{Final Remarks}
Let us summarize our work. Once again we have advocared a key point
for understanding the physical content of gauge theories, that is, the identification of field degrees of freedom with observable quantities.
We have showed  that the static potential profile obtained from both 
a gauge theory which includes a light massive vector field interacting with the familiar photon $U(1)_{QED}$ via a Chern-Simons-like coupling
and axionic electrodynamics models are quite different. This means that the two theories are not equivalent. As it was shown in \cite{GaeteGuen}, axionic electrodynamics has a different structure which is reflected in a confining piece, which is not present in the gauge theory which includes a light massive vector field interacting with the familiar photon $U(1)_{QED}$ via a Chern-Simons-like coupling. However, our result is analogous
to that encountered in the coupling between the familiar photon
$U(1)_{QED}$ and a second massive gauge field living in the
so-called hidden-sector $U(1)_h$, inside a superconducting box. \\

\section{Acknowledgments}

One of us (PG) wants to thank the Field Theory Group of the CBPF and
the Physics Department  of the Universit\`a di Trieste for
hospitality. This work was supported in part by Fondecyt (Chile)
grant 1080260 (PG). (J.H-N) expresses his gratitude to CNPq.

\end{document}